\documentclass[12pt]{article}
\usepackage{bm}% bold math
\begin{document}

\title{Nonzero $\theta_{13}$ and Neutrino Masses from Modified TBM}
\author{\bf{Asan Damanik}\footnote{E-mal: d.asan@lycos.com}\\Faculty of Science and Technology\\Sanata Dharma University\\Kampus III USD Paingan Maguwoharjo Sleman Yogyakarta\\Indonesia}
\date{}

\maketitle

\abstract{In order to accommodate nonzero and relatively large of mixing angle $\theta_{13}$, we modified the tribimaximal mixing(TBM) matrix by introducing a simple perturbation matrix to perturb TBM matrix.  The modified TBM can reproduce nonzero mixing angle $\theta_{13}=7.9^{0}$ which is in agreement with the present  experimental results.  By imposing two zeros texture into the obtained neutrino mass matrix from modified TBM, we then have the neutrino mass spectrum in normal hierarchy.  Some phenomenological implications are also discussed.}

\section{Introduction}
There are three types of the well-known neutrino mixing matrices; tribimaximal (TBM), bimaximal (BM), and democratic DC).  These three neutrino mixing matrices patterns predict the  mixing angle $\theta_{13}=0$.  Recently, the evidence of nonzero $\theta_{13}$ due to the achievement of experimental methods and tools, the assumption that the value of mixing angle $\theta_{13}$ is very small and tend to zero must be corrected or even ruled out.  Concerning with the well-known mixing matrix, especially tribimaximal neutrino mixing matrix, Ishimori and Ma \cite{Ishimori} stated explicitly that the tribimaximal mixing matrix may be dead due to the experimental fact that mixing angle $\theta_{13}$  is not zero.  The nonzero and relatively large mixing angle $\theta_{13}$ have already been reported by MINOS \cite{Minos}, Double Chooz \cite{Double}, T2K \cite{T2K},  Daya Bay  \cite{Daya}, and RENO \cite{RENO} collaborations.

The evidence of nonzero and relatively large $\theta_{13}$ as reported by many collaborations, several authors have already proposed some methods and models in order to explain the existence of nonzero $\theta_{13}$.  The simple way to accommodate a nonzero $\theta_{13}$ is to modify the neutrino mixing matrix by introducing a perturbation matrix into known mixing matrix such that it can produces a nonzero $\theta_{13}$ \cite{He11, Damanik, Brahmachari}, breaking the scaling ansatz \cite{Biswajit},  and the other is to build the model by using some discrete symmetries \cite{Cao, Luca, Ge1, Ge2}.

In this paper we modify TBM mixing matrix by introducung a simple perturbation matrix and calculate the mixing angle $\theta_{13}$ by using the advantages of the mixing angles $\theta_{21}$ and $\theta_{32}$ from the experimental results.  The modofied TBM is used to construct the neutrino mass matrix and we evaluate the neutrino mass and its hierarchy.  This paper is organized as follow: in section 2, we modify tribimaximal mixing matrix by introducing a simple perturbation matrix.  In section 3, we determine the neutrino mass spectrum from modified tribimaximal mixing matrix.  Finally, section 4 is devoted to conclusion.

\section{Nonzero $\theta_{13}$ from the modified tribimaximal mixing matrix}

The TBM mixing matrix existence is due to the experimental facts that mixing of flavors do exist in the leptonic sector especially in  neutrino sector as well as in the quarks sector. The neutrino eigenstates in flavor basis ($\nu_{e}, \nu_{\mu}, \nu_{\tau}$) relate to the eigenstates of neutrino in mass basis ($\nu_{1}, \nu_{2}, \nu_{3}$) as follow:
\begin{eqnarray}
\nu_{i}=V_{ij}\nu_{j},
\end{eqnarray}
where $V_{ij} ( i = e, \mu, \tau; j = 1,2,3)$ are the elements of neutrino mixing matrix.  The mixing matrix $V$ can be parameterized as follow:
\begin{eqnarray}
V=\bordermatrix{& & &\cr
&c_{12}c_{13} &s_{12}c_{13} &s_{13}e^{-i\phi}\cr
&-s_{12}c_{23}-c_{12}s_{23}s_{13}e^{i\phi} &c_{12}c_{23}-s_{12} s_{23}s_{13}e^{i\phi}&s_{23}c_{13}\cr
&s_{12}s_{23}-c_{12}c_{23}s_{13}e^{i\phi} &-c_{12}s_{23}-s_{12}c_{23}s_{13}e^{i\phi} &c_{23}c_{13}}
 \label{V}
\end{eqnarray}
where $c_{ij}$ is the $\cos\theta_{ij}$, $s_{ij}$ is the $\sin\theta_{ij}$, and $\theta_{ij}$ are the mixing angles.

One of the well-known neutrino mixing matrix ($V$) is the tribimaximal neutrino mixing matrix ($V_{TBM}$) which given by \cite{Harrison, Harrisona, Xing, Harrisonb, Harrisonc, He}:
\begin{eqnarray}
V_{TBM}=\bordermatrix{& & &\cr
&\sqrt{\frac{2}{3}} &\frac{1}{\sqrt{3}} &0\cr
&-\frac{1}{\sqrt{6}} &\frac{1}{\sqrt{3}} &\frac{1}{\sqrt{2}}\cr
&-\frac{1}{\sqrt{6}} &\frac{1}{\sqrt{3}} &-\frac{1}{\sqrt{2}}}.
 \label{tb}
\end{eqnarray}
As one can see from Eq. (\ref{tb}) that the entry $V_{e3}=0$ which imply that the mixing angle $\theta_{13}$ must be zero in the tribimaximal mixing matrix.   However, the latest result from long baseline neutrino oscillation experiment T2K indicates that $\theta_{13}$ is relatively large.  For a vanishing Dirac CP-violating phase, the T2K collaboration reported that the values of $\theta_{13}$  are \cite{T2K}:
\begin{equation}
5.0^{o}\leq\theta_{13}\leq 16.0^{o}, {\rm and}~5.8^{o}\leq\theta_{13}\leq 17.8^{o},
\end{equation}
for neutrino mass in norma (NH)l and inverted (IH) hierarchies respectively,  and the current combined world data\cite{Gonzales-Carcia}-\cite{Fogli}:
\begin{equation}
\Delta m_{21}^{2}=7.59\pm0.20 (_{-0.69}^{+0.61}) \times 10^{-5}~\rm{eV^{2}},\label{21}
\end{equation}
\begin{equation}
\Delta m_{32}^{2}=2.46\pm0.12(\pm0.37) \times 10^{-3}~\rm{eV^{2}},~\rm(for~ NH)\label{32}
\end{equation}
\begin{equation}
\Delta m_{32}^{2}=-2.36\pm0.11(\pm0.37) \times 10^{-3}~\rm{eV^{2}},~\rm(for~ IH)\label{321}
\end{equation}
\begin{equation}
\theta_{12}=34.5\pm1.0 (_{-2.8}^{3.2})^{o},~~\theta_{23}=42.8_{-2.9}^{+4.5}(_{-7.3}^{+10.7})^{o},~~\theta_{13}=5.1_{-3.3}^{+3.0}(\leq 12.0)^{o},
 \label{GD}
\end{equation}
at $1\sigma~(3\sigma)$ level.  The latest experimental result on $\theta_{13}$ is reported by Daya Bay Collaboration which gives \cite{Daya}:
\begin{equation}
\sin^{2}2\theta_{13}=0.092\pm 0.016 (\rm{stat}.)\pm 0.005 (\rm{syst.}),
\end{equation}
and RENO Collaboration reported that \cite{RENO}:
\begin{equation}
\sin^{2}2\theta_{13}=0.113\pm 0.013 (\rm{stat.})\pm 0.014 (\rm{syst.}).
\end{equation}

Modification of neutrino mixing matrix, by introducing  a perturbation matrices into neutrino mixing matrices in Eq. (\ref{tb}), is the easiest way to obtain the nonzero $\theta_{13}$.  The value of $\theta_{13}$ can be obtained in some parameters that can be fitted from experimental results.  In this paper, the modified neutrino mixing matrices to be considered are given by:
\begin{equation}
V_{{\rm TBM}}^{'}=V_{{\rm TBM}}V_{y},\label{Modi1}
\end{equation}
where $V_{y}$ is the perturbation matrices to the neutrino mixing matrices.  We take the form of the perturbation matrices as follow:
\begin{equation}
V_{y}=\bordermatrix{& & &\cr
&1 &0 &0\cr
&0 &c_{y} &s_{y}\cr
&0 &-s_{y} &c_{y}\cr}.
 \label{xy}
\end{equation}
where $c_{y}$ is the $\cos{y}$, and $s_{y}$ is the $\sin{y}$.

By inserting Eqs. (\ref{tb}) and (\ref{xy}) into Eqs. (\ref{Modi1}), we then have the modified neutrino mixing matrices as follow:
\begin{equation}
V_{{\rm TB}}^{'}=\bordermatrix{& & &\cr
&\frac{\sqrt{6}}{3} &\frac{\sqrt{3}}{3}c_{y} &\frac{\sqrt{3}}{3}s_{y}\cr
&-\frac{\sqrt{6}}{6} &\frac{\sqrt{3}}{3}c_{y}-\frac{\sqrt{2}}{2}s_{y} &\frac{\sqrt{3}}{3}s_{y}+\frac{\sqrt{2}}{2}c_{y}\cr
&-\frac{\sqrt{6}}{6} &\frac{\sqrt{3}}{3}c_{y}+\frac{\sqrt{2}}{2}s_{y} &\frac{\sqrt{3}}{3}s_{y}-\frac{\sqrt{2}}{2}c_{y}},\label{Mo1}
\end{equation}
By comparing Eqs. (\ref{Mo1}) with the neutrino mixing in standard parameterization form as shown in Eq. (\ref{V}) with $\varphi=0$, then we obtain:
\begin{equation}
\tan\theta_{12}=\left|\frac{\sqrt{2}c_{y}}{2}\right|,~~
\tan\theta_{23}=\left|\frac{\frac{\sqrt{3}}{3}s_{y}+\frac{\sqrt{2}}{2}c_{y}}{\frac{\sqrt{3}}{3}s_{y}-\frac{\sqrt{2}}{2}c_{y}}\right|,~~
\sin\theta_{13}=\left|\frac{\sqrt{3}}{3}s_{y}\right|.
 \label{1}
\end{equation}

From Eq. (\ref{1}) it is apparent that for $y\rightarrow 0$, the value of $\tan\theta_{12}\rightarrow \sqrt{2}/2$  and $\tan\theta_{23}\rightarrow 1$ which imply that $\theta_{12}\rightarrow 35.264^{o}$ and $\theta_{23}\rightarrow 45^{o}$.   From Eq. (\ref{1}), one can see that it is possible to determine the value $y$ and therefore the value of $\theta_{13}$ by using the experimental values of $\theta_{12}$ and $\theta_{23}$ in Eq. (\ref{GD}).

By inserting the experimental values of $\theta_{12}$ and $\theta_{23}$ in Eq. (\ref{GD}) into Eq. (\ref{1}), we obtain the relations:
\begin{equation}
c_{y}=-0.03167630078 s_{y},\label{c1}
\end{equation}
when we use $\theta_{23}$, and
\begin{equation}
c_{y}=0.9713265692,\label{c2}
\end{equation}
when we use $\theta_{12}$.  From both Eqs. (\ref{c1}) and (\ref{c2}), we can see that the realistic value for $c_{y}$ is the value $c_{y}$ in Eq. (\ref{c2}) that is $y=13.7537^{o}$.  It means that in this modification scenario, only the experimental mixing angle $\theta_{12}$ related to the mixing angle $\theta_{13}$.  From Eq. (\ref{c2}), we have:
\begin{equation}
\sin\theta_{13}=0.137265,
\end{equation}
that imply the mixing angle $\theta_{13}=7.89^{o}$ which is in agreement with the T2K \cite{T2K} and Daya Bay experimental results \cite{Daya}.

\section{Neutrino masses from modified tribimaximal mixing matrix}

We construct the neutrino mass matrix $M_{\nu}$ in flavor eigenstates basis (where the charged lepton mass matrix is diagonal).  In this basis, the neutrino mass matrix can be diagonalized by a unitary matrix $V$ as follow:
\begin{equation}
M_{\nu}=VMV^{T},\label{aa}
\end{equation}
where the diagonal neutrino mass matrix $M=diag(m_{1},m_{2},m_{3})$.

If we put $V$ is the modified neutrino mixing matrix in Eq. (\ref{Mo1}), then Eq. (\ref{aa}) gives the neutrino mass matrix:
\begin{equation}
M_{\nu}=\bordermatrix{& & &\cr
&A &B &C\cr
&B &D &E\cr
& C&E &F}=\bordermatrix{& & &\cr
&(M_{\nu})_{11} &(M_{\nu})_{12} &(M_{\nu})_{13}\cr
&(M_{\nu})_{21} &(M_{\nu})_{22} &(M_{\nu})_{23}\cr
&(M_{\nu})_{31} &(M_{\nu})_{32} &(M_{\nu})_{33}},\label{Mv}
\end{equation}
where:
\begin{equation}
(M_{\nu})_{11}=\frac{2m_{1}}{3}+\frac{m_{2}}{3}c_{y}^{2}+\frac{m_{3}}{3}s_{y}^{2},\label{M11}
\end{equation}
\begin{equation}
(M_{\nu})_{12}=(M_{\nu})_{21}=-\frac{m_{1}}{3}+m_{2}\left(\frac{1}{3}c_{y}^{2}-\frac{\sqrt{6}}{6}c_{y}s_{y}\right)+m_{3}\left(\frac{1}{3}s_{y}^{2}+\frac{\sqrt{6}}{6}s_{y}c_{y}\right),\label{M12}
\end{equation}
\begin{equation}
(M_{\nu})_{13}=(M_{\nu})_{31}=-\frac{m_{1}}{3}+m_{2}\left(\frac{1}{3}c_{y}^{2}+\frac{\sqrt{6}}{6}c_{y}s_{y}\right)+m_{3}\left(\frac{1}{3}s_{y}^{2}-\frac{\sqrt{6}}{6}s_{y}c_{y}\right),\label{M13}
\end{equation}
\begin{equation}
(M_{\nu})_{22}=\frac{m_{1}}{6}+m_{2}\left(\frac{\sqrt{3}}{3}c_{y}-\frac{\sqrt{2}}{2}s_{y}\right)^{2}+m_{3}\left(\frac{\sqrt{3}}{3}s_{y}+\frac{\sqrt{2}}{2}c_{y}\right)^{2},\label{M14}
\end{equation}
\begin{equation}
(M_{\nu})_{23}=(M_{\nu})_{32}=\frac{m_{1}}{6}+m_{2}\left(\frac{1}{3}c_{y}^{2}-\frac{1}{2}s_{y}^{2}\right)+m_{3}\left(\frac{1}{3}s_{y}^{2}-\frac{1}{2}c_{y}^{2}\right),\label{M15}
\end{equation}
\begin{equation}
(M_{\nu})_{33}=\frac{m_{1}}{6}+m_{2}\left(\frac{\sqrt{3}}{3}c_{y}+\frac{\sqrt{2}}{2}s_{y}\right)^{2}+m_{3}\left(\frac{\sqrt{3}}{3}s_{y}-\frac{\sqrt{2}}{2}c_{y}\right)^{2}.\label{M16}
\end{equation}

To simplify the problem such that we can determine the neutrino masses, which can correctly predict the neutrino mass spectrum, we impose texture zero into neutrino mass matrix in Eq. (\ref{Mv}).  Texture zero of neutrino mass matrix indicates the existence of additional symmetries beyond the Standard Model Particle Physics \cite {Damanik2007, Fritzsch2011}.   By imposing some possibilities texture zero into Eq. (\ref{Mv}), we then find that only one texture zero: $(M_{\nu})_{11}=(M_{\nu})_{13}=0$ can correctly predict the nuetrino mass spectrum.  From this texture zero pattern, we have:
\begin{equation}
m_{2}=-1.400444385 m_{1},~m_{3}=-12.00741191 m_{1},\label{mm}
\end{equation}
that implies that the neutrino mass hierarchy is normal hierarchy: $\left|m_{1}\right|<\left|m_{2}\right|<\left|m_{3}\right|$.

If we use the experimental value of the solar neutrino squared-mass difference ($\Delta m_{21}^{2}$) in Eq. (\ref{21}) to determine the neutrino masses in Eq. (\ref{mm}), then we have:
\begin{equation}
m_{1}=0.00888595~{\rm eV},~m_{2}=0.01244428~{\rm eV},~m_{3}=0.10669729~{\rm eV}. \label{m}
\end{equation}
The obtained neutrino masses in Eq. (\ref{m}) cannot give correctly the squared-mass difference for atmospheric neutrino ($\Delta m_{32}^{2}$) in Eq. (\ref{32}).  Conversely, if we use the experimental value of $\Delta m_{32}^{2}$ in Eq. (\ref{32}) to determine  the value of neutrino masses in Eq. (\ref{mm}), then the obtained neutrino masses cannot correctly predict the squared-mass difference for solar neutrino in Eq. (\ref{21}).

\section{Conclusion}
By introducing a simple perturbation matrix into tribimaximal mixing matrix,  we then have the modified tribimaximal neutrino mixing matrix that can give nonzero $\theta_{13}=7.89^{o}$ which is in agreement with the present experimental results.  The neutrino mass matrix from the modified tribimaximal neutrino mixing matrix with two zeros texture  predict the neutrino mass spectrum in normal hierarchy: $\left|m_{1}\right|<\left|m_{2}\right|<\left|m_{3}\right|$.  If we use the solar neutrino squared-mass difference to determine the values of neutrino masses, then we cannot have the correct value for the atmospheric squared-mass difference.  Conversely, if we use the experimental valeu of the squared-mass difference to determine the neutrino masses, then we cannot have the correct value for the solar neutrino squared-mass difference.

\end{document}